\documentstyle[12pt,prd,aps,epsf]{revtex}
\begin{document}
\baselineskip=20pt
\draft
\title{Neutrino conversions in cosmological gamma-ray burst fireballs}
\author{H. Athar}
\address{Department of Physics, Tokyo Metropolitan University, Minami-Osawa 
 1-1, Hachioji-Shi, \\ 
 Tokyo 192-0397, Japan; E-mail: athar@phys.metro-u.ac.jp}
\maketitle
\begin{abstract} 
\tightenlines

	We study neutrino conversions in a  recently envisaged source 
of high energy neutrinos ($E {\buildrel > \over {_{\sim}}} \, 10^{6}$ GeV),  
that is, in the vicinity of cosmological Gamma-Ray Burst fireballs
 (GRB). We consider the effects of flavor and spin-flavor conversions 
and point out that in both situations, a some what higher than estimated
 high energy tau neutrino flux from GRBs is expected in new km$^{2}$ surface 
 area under water/ice neutrino telescopes.

\end{abstract}


\section{Introduction}
	Recently, cosmological fireballs are suggested as a 
possible
production site for gamma-ray bursts as well as the high energy 
 ($E {\buildrel > \over {_{\sim}}} \, 10^{6}$ GeV)  neutrinos \cite{WB}. 
Although,
 the origin of these cosmological Gamma-Ray Burst fireballs (GRB) 
 is not 
yet understood, the observations suggest that generically 
a very compact source of  linear scale $\sim \, 10^{7}$ cm  through 
 internal or/and external shock propagation produces these 
gamma-ray bursts (as well as the burst of high energy neutrinos) \cite{P}.
 Typically, this compact source is hypothesized to be formed possibly
due to the merging of binary neutron stars or due to collapse of a 
supermassive star. 

The main source of high energy tau neutrinos in GRBs is the production 
 and decay of $D^{\pm}_{S}$. The production of $D^{\pm}_{S}$ may be  
 through $p\gamma $ and/or  through $pp$ collisions.
	In \cite{PX}, the $\nu_{e}$ and $\nu_{\mu}$  flux is 
estimated in $pp$
collisions, whereas in \cite{WB}, the $\nu_{e}$ and $\nu_{\mu}$ flux is 
estimated in $p\gamma$ collisions for GRBs. 
In $pp$ collisions, the flux of $\nu_{\tau}$ may be obtained through the main 
process of
$p+p\rightarrow D^{+}_{S}+X$. The $D^{+}_{S}$ decays into 
$\tau^{+}$ 
lepton and $\nu_{\tau}$ with a branching ratio of $\sim \, 3\%$. This 
$\tau^{+}$ lepton
further decays into $\nu_{\tau}$.
The cross-section for  $D^{+}_{S}$ production, which is main source of 
$\nu_{\tau}$'s,  
is $\sim $ 4 orders of magnitude lower than that of $\pi^{+}$ and/or $\pi^{-}$ 
 which subsequently produces 
$\nu_{e}$ and $\nu_{\mu}$. The branching ratio for 
$\nu_{e}$ and/or
$\nu_{\mu}$ production is higher up to an order of magnitude 
 than that for $\nu_{\tau}$ production (through $D^{\pm}_{S}$). These imply 
that the 
$\nu_{\tau}$ flux in $pp$ collisions is suppressed up to 5 orders of 
magnitude 
relative to corresponding $\nu_{e}$ and/or $\nu_{\mu}$ fluxes. 
 In $p\gamma$ collisions, 
the main
process for the production of $\nu_{\tau}$ may be 
  $p+\gamma \rightarrow D^{+}_{S}+
\Lambda^{0}+\bar{D}^{0}$ with similar relevant branching ratios and 
 corresponding suppression for cross-section  values. 
 Here the corresponding main source for $\nu_{e}$ and $\nu_{\mu}$ 
 production is 
 $p+\gamma \, \rightarrow \Delta^{+}\rightarrow \pi^{+}+n$. 
Therefore, in $p\gamma $ collisions, the $\nu_{\tau}$ flux is also 
 suppressed up to 5 orders 
 of magnitude relative to $\nu_{e}$ and/or $\nu_{\mu}$ flux. Thus, in both 
type of collisions, including the relevant kinematics, the  
 intrinsic $\nu_{\tau}(\bar{\nu}_{\tau})$ flux 
is estimated to be rather small relative to $\nu_{e} (\bar{\nu}_{e})$ 
 and/or $\nu_{\mu} (\bar{\nu}_{\mu})$ 
  fluxes from GRBs, typically being, $F^{0}_{\tau}
 /F^{0}_{e,\mu}\, <10^{-5}$ \cite{VZA}.

	In this paper, we consider the possibility of obtaining higher 
$\nu_{\tau}$ flux, that is, 
 $F^{0}_{\tau}/F^{0}_{e,\mu}\, >\, 10^{-5}$, 
 from GRBs through neutrino 
conversions
as compared to no conversion situation.
	The present study is particularly useful as the new under ice/water 
 \v{C}erenkov light neutrino
telescopes namely AMANDA and Baikal (also the NESTOR and ANTARES)
 will have energy, angle 
and flavor resolution for high energy neutrinos originating at cosmological 
distances \cite{M}. Recently, there are several discussions concerning the 
signatures 
of a possible neutrino burst from GRBs correlated in time and 
angle 
\cite{W}. In particular, there is a suggestion of measuring $\nu_{\tau}$ 
flux from 
cosmologically distant  sources through a double shower 
 (bang) event \cite{L} or 
 through a small pile up of up ward going $\mu$-like  events near 
 (10$^{4}-10^{5}$) GeV \cite{HS}.

	The plan of this paper is as follows. 
 In Sect. II, we briefly describe  the matter density
and magnetic field in the vicinity of GRBs. In Sect. III, we discuss in 
some detail, the range of neutrino mixing parameters that may give rise to 
relatively large precession/conversion probabilities resulting from 
 neutrino flavor/spin-flavor conversions.
 In Sect. IV, we give  estimates for separable but contained  double shower
 event rates induced by high energy $\nu_{\tau}$'s originating from GRBs 
  without/with conversions for km$^{2}$ surface area under water/ice 
 neutrino telescopes for illustrative purposes  
 and finally in Sect. V, we summarize our results.

\section{matter density and magnetic field in the vicinity of GRB}

According to \cite{WB}, 
 the isotropic high energy neutrino production may take place in the 
 vicinity of
$r_{p}\, \sim \, \Gamma^{2}c\Delta t$. Here $\Gamma $ is the Lorentz 
factor (typically $\Gamma \, \sim \, 300$) and $\Delta t$ is the observed
GRB variability time scale (typically $\Delta t\, \sim \, 1$ ms).   
  In the vicinity of $r_{p}$, the fireball matter density is 
$\rho \, \sim \, 10^{-13}$ g cm$^{-3}$ \cite{WB}. 
 In these models, the typical distance traversed by neutrinos 
may be
taken as, $\Delta r\,  {\buildrel < \over {_{\sim}}}\, (10^{-4}-1)$ pc, 
 in the vicinity of 
 GRB, where 1 pc $\sim \, 3\times 10^{18}$ cm.
	Matter effects on neutrino oscillations are
relevant if $G_{F}\rho /m_{N}\, \sim \, \delta m^{2}/2E$. Using $\rho$ 
from Ref. \cite{WB}, it follows  that matter effects are 
absent for $\delta m^{2}\, {\buildrel > \over {_{\sim}}}
 \, {\cal O}(10^{-10})$ eV$^{2}$. Matter effects 
 due to coherent forward scattering of neutrinos off the
  background for high energy neutrinos originating from GRBs
are not expected to be important in the neutrino 
production regions around GRBs and will not be further discuss here.

Taking the observed duration of the typical gamma-ray burst as, $\Delta t\, 
  {\buildrel < \over {_{\sim}}}
 \, 1$ ms, we obtain the mass of the source as, $M_{GRB}\, 
 {\buildrel < \over {_{\sim}}}\, \Delta t/G_{N}$, where $G_{N}$ is the 
 gravitational constant. For the relatively shorter observed duration of  
 gamma-ray burst from a typical GRB,
 $\Delta t\, \sim \, 0.2 $ ms, implying $M_{GRB}\, \sim \, 40\, M_{\odot}$
 (where $M_{\odot}\, \sim \, 2\times 10^{33}$ g is solar mass). 
We use $M_{GRB}\, \sim \, 2\times 10^{2}\,M_{\odot}$ in our estimates.  

	The magnetic field in the vicinity of a GRB is obtained 
by considering the equipartition arguments \cite{WB}.
We use the following profile of magnetic field, $B_{GRB}$,  
as an example, for $r\, > \, r_{p}$ \cite{mag}
\begin{equation}
 B_{GRB}(r)\, \simeq \, B_{0}\left(\frac{r_{p}}{r}\right)^{2},
\label{bprofile}
\end{equation}
where $B_{0}\, \sim \, L^{1/2}c^{-1/2}(r_{p}\Gamma)^{-1}$ with $L$ being the 
total wind luminosity (typically $L\, \sim \, 10^{51}$ erg s$^{-1}$).

\section{Neutrino conversions in GRB}

\subsection{Flavor oscillations}

	In the framework of three flavor analysis, the flavor precession
probability from $\alpha $ to $\beta $ neutrino flavor is \cite{book} 

\begin{equation}
 P(\nu_{\alpha}\rightarrow \nu_{\beta})  \equiv P_{\alpha \beta}  = 
 \sum^{3}_{i=1}|U_{\alpha i}|^{2}|U_{\beta i}|^{2}
 +\sum_{i \neq j} U_{\alpha i}U^{\ast}_{\beta i}
                        U^{\ast}_{\alpha j}U_{\beta j}
                        \cos\left(\frac{2\pi L}{l_{ij}}\right),
\label{3flvp}
\end{equation}
where $\alpha, \beta = e, \mu, $ or $\tau $. $U$ is the 3$\times $3 MNS mixing 
matrix and can be obtained in usual notation through 
\begin{equation}
 U\, \equiv R_{23}(\theta_{1})
 \mbox{diag}(e^{-i\delta /2},1,e^{i\delta /2})  
 R_{31}(\theta_{2}) \mbox{diag}(e^{i\delta /2},1,e^{-i\delta /2})
       R_{12}(\theta_{3}),
\label{defU} 
\end{equation}
thus coinciding with the standard form given by the Particle Data Group 
\cite{pdg}. In Eq. (\ref{defU}), $l_{ij}\simeq 4\pi E/
\delta m^{2}_{ij}$ with $\delta m^{2}_{ij} \equiv |m^{2}_{i}-m^{2}_{j}|$ and 
$L$ is the distance between the source and the detector. For simplicity, 
 we will assume here a vanishing value for $\theta_{31}$ and CP violating 
 phase $\delta $ in $U$. 

	At present, the atmospheric muon and solar electron neutrino deficits 
 can be explained with 
oscillations among three active neutrinos \cite{RECENT}. For this, 
 typically, $\delta m^{2}
\sim {\cal O}(10^{-3})$ eV$^{2}$ and $\sin^{2}2\theta \sim {\cal O}(1)$ for 
the explanation of atmospheric muon neutrino deficit, 
whereas for the explanation of solar electron neutrino deficit, we may have  
$\delta m^{2} \sim {\cal O}(10^{-10})$ eV$^{2}$ and 
$\sin^{2}2\theta \sim {\cal O}(1)$ [just so] or $\delta m^{2}
\sim {\cal O}(10^{-5})$ eV$^{2}$ and $\sin^{2}2\theta \sim {\cal O}(10^{-2})$
[SMA (MSW)] or $\delta m^{2}
\sim {\cal O}(10^{-5})$ eV$^{2}$ and $\sin^{2}2\theta \sim {\cal O}(1)$ 
[LMA (MSW)]. The present status of data thus permits multiple oscillation 
 solutions to solar neutrino deficit. We  intend to discuss here
implications of these mixings for high energy cosmic neutrino propagation.

	In the above explanations, the total range of $\delta m^{2}$ is 
$10^{-10}\leq \delta m^{2}/$ eV$^{2} \leq 10^{-3}$ irrespective of neutrino 
 flavor. The typical energy span relevant for possible flavor 
 identification for high
energy cosmic neutrinos is $2\times 10^{6}\leq E/$GeV$\leq 2\times 10^{7}$ 
 (see Sect. IV). 
 Taking a typical distance between the GRB and our galaxy as 
 $L \sim 1000$ Mpc, we note that $\cos $ term in Eq. (\ref{3flvp})
 vanishes and so Eq. (\ref{3flvp}) reduces to 
 
\begin{equation}
   P_{\alpha \beta}  \simeq  
   \sum^{3}_{i=1}|U_{\alpha i}|^{2}|U_{\beta i}|^{2}.
\label{preduced}
\end{equation}

It is assumed here that no relatively dense objects exist between the GRB and 
the earth so as to effect significantly this oscillations pattern.
 Since $P_{\alpha \beta }$ in above Eq. is 
 symmetric under the exchange of $\alpha $ and
$\beta $ indices implying that essentially no $T$ (or $CP$) violation effects
 arise in 
neutrino vacuum flavor oscillations for high energy cosmic neutrinos 
\cite{cabibbo}.
 
	Let us denote by $F^{0}_{\alpha }$, the intrinsic neutrino fluxes. 
From the discussion in the previous Sect., it follows that 
$F^{0}_{e} : F^{0}_{\mu} :F^{0}_{\tau} = 1 :2 : < 10^{-5}$. For simplicity, 
we take these ratios as 1 : 2 : 0. In order to estimate the final flux  ratios 
on earth for high energy cosmic neutrinos originating from cosmologically 
distant GRBs, let us  introduce a 3$\times $3 matrix of vacuum flavor 
precession probabilities such that 

\begin{equation}
 F_{\alpha} = \sum_{\beta}P_{\alpha \beta} F^{0}_{\beta},
\label{falpha}
\end{equation}
where  the unitarity conditions for $ P_{\alpha \beta} $ read as

\begin{eqnarray}
 P_{ee} + P_{e\mu } + P_{e\tau } &  = & 1,\nonumber \\
 P_{e\mu} + P_{\mu \mu } + P_{\mu \tau }&  = & 1,\nonumber \\
 P_{e\tau }+ P_{\mu \tau } + P_{\tau \tau } &  = & 1.
\label{unitarity} 
\end{eqnarray}

The explicit form for the matrix $P $ in case of 
just so flavor oscillations as solution to solar neutrino problem along with
the solution to atmospheric neutrino deficit in terms of $\nu_{\mu}$ to 
$\nu_{\tau}$ oscillations with maximal depth is

\begin{equation}
 P  = \left( \begin{array}{ccc}
                             1/2 & 1/4 & 1/4 \\
                             1/4 & 3/8 & 3/8 \\
                             1/4 & 3/8 & 3/8 
                            \end{array}
                      \right).
\label{justso}
\end{equation}
Using Eq. (\ref{justso}) and Eq. (\ref{falpha}), we note that 
 $F_{e}: F_{\mu }: F_{\tau } = 1:
1: 1$ at the level of $F^{0}_{e}$. Also, Eq. (\ref{unitarity}) is 
 satisfied. The same final 
flux ratio is obtained in remaining two cases for which the corresponding 
 $P$ matrices are [for SMA (MSW)]

\begin{equation}
 P  = \left( \begin{array}{ccc}
                             1 & 0 & 0 \\
                             0 & 1/2 & 1/2 \\
                             0 & 1/2 & 1/2 
                            \end{array}
                      \right),
\label{sma}
\end{equation}
whereas in case of LMA (MSW),

\begin{equation}
  P  = \left( \begin{array}{ccc}
                             5/8  & 3/16 & 3/16 \\
                             3/16 & 13/32 & 13/32 \\
                             3/16 & 13/32 & 13/32 
                            \end{array}
                      \right).
\label{lma}
\end{equation}

Thus, independent of the 
oscillation solution for solar neutrino problem, we have  
 $F_{e}: F_{\mu }: F_{\tau } = 1: 1: 1$. Importantly, these ratios do
not depend on neutrino energy or $\delta m^{2}$ at least in the relevant 
energy interval.

	Summarizing, although intrinsically the high energy cosmic tau 
neutrino flux is negligibally small however because of vacuum flavor 
oscillations it becomes comparable to $\nu_{e}$ flux thus providing some 
prospects for its possible detection.

\subsection{Spin-flavor oscillations}

	We consider here an example of a possibility that may lead to an 
energy dependence and/or change in the above mentioned final flux ratios. 
 We consider spin-flavor oscillations 
resulting from an interplay of possible Violation of 
 Equivalence Principle (VEP) parameterized by $\Delta f$ and the 
 magnetic field in the vicinity of GRBs.  
 We obtain the range of neutrino mixing parameters 
 giving $F_{\tau}/F_{e,\mu} >  \, 10^{-5}$. 
 The possibility of VEP  arises from the realization 
 that different flavors of neutrinos may 
couple differently to gravity \cite{G}.

	Consider a system of two mixed neutrinos ($\bar{\nu}_{e}$ 
and $\nu_{\tau}$) for simplicity. The difference of diagonal elements of 
 the $2\times 2$ effective Hamiltonian describing the dynamics of the 
 mixed system of these 
 oscillating neutrinos in the basis 
 $\psi^{T}\, =\, (\bar{\nu}_{e}, \nu_{\tau})$ for vanishing vacuum and 
gravity mixing angles is \cite{AK}
\begin{equation}
 \Delta H\, =\, \delta  - V_{G}, 
\label{deltaH}
\end{equation}
whereas each of the off diagonal elements is $\mu B$ ($\mu $ is magnitude of 
 neutrino magnetic moment). In Eq. (\ref{deltaH}), $\delta \, =\, 
\delta m^{2}/2E$, where $\delta m^{2}\, =\, m^{2}
(\nu_{\tau})-m^{2}(\bar{\nu}_{e}) \, >\, 0$.  
Here $V_{G}$ is the effective potential felt by the neutrinos at a 
distance $r$ 
from a gravitational source of mass $M$ due to VEP and is given by \cite{G}
\begin{equation}
 V_{G} \, \equiv \, \Delta f \phi(r)E,
\label{defVg}
\end{equation}
where $\Delta f \, =\, f_{3}-f_{1}$ is a measure of the 
degree
of VEP and $\phi(r)\, = \, G_{N}Mr^{-1}$ is the gravitational potential 
in the 
Keplerian approximation. 
 In Eq. (\ref{defVg}), $f_{3}G_{N}$ and $f_{1}G_{N}$ are 
 respectively the gravitational couplings of $\nu_{\tau}$ and 
 $\bar{\nu}_{e}$, such that $f_{1}\, \neq \, f_{3}$.
 We assume here same gravitational couplings for $\nu $ and $\bar{\nu}$ for 
 simplicity. This implies that the sign of $V_{G}$ is same for 
 $\bar{\nu}_{e}\rightarrow \nu_{\tau}$ and 
 $\nu_{e}\rightarrow \bar{\nu}_{\tau}$ conversions \cite{AK}. 
 If this is not the case then the sign of $V_{G}$ will be different for the 
 two conversions and only 
 one of the two conversions can occur. This may lead to a change in expected 
 $\nu_{e}/\bar{\nu}_{e}$ ratio. 
 We briefly comment on the possibility of empirical realization of this 
 situation in next Sect.. 
 
There are at least three relevant $\phi (r)$'s that need to be 
 considered \cite{MS}.
 The effect of $\phi (r)$ due to supercluster named Great Attractor 
 with $\phi_{SC} (r)$ in the vicinity of GRB;
 $\phi (r)$ due to GRB itself, which is,  
 $\phi_{GRB} (r)$, in the vicinity of GRB and the  
 galactic gravitational potential, which is,  $\phi_{G}(r)$. 
 Therefore, we use, 
 $\phi(r)\, \equiv \, \phi_{SC}(r)+\phi_{GRB}(r)+\phi_{G}(r)$.
However, $\phi_{G}(r)\, \ll \, \phi_{SC}(r),\, \phi_{GRB}(r)$ in the 
 vicinity of GRB. Thus, 
$\phi(r)\, \simeq \, \phi_{SC}(r)+\phi_{GRB}(r)$. If the neutrino 
production region  $r_{p}$ is  ${\buildrel < \over {_{\sim}}} 10^{13}$ cm then
at $r\, \sim \, r_{p}$, 
 we have $\phi_{GRB}(r)\, >\, \phi_{SC}(r)$. At $r\, \sim \,
6\times 10^{13}$ cm, $\phi_{GRB}(r)\, \sim \, \phi_{SC}(r)$ and for 
$r\,  {\buildrel > \over {_{\sim}}}\, 10^{14}$ cm, $\phi_{GRB}(r)\, <\,
\phi_{SC}(r)$. If the supercluster is a fake object then $\phi(r)\, \simeq \,
\phi_{GRB}(r)$. 
 Here we assume the smallness of the effect of $\phi (r)$ due to an 
 Active Galactic Nucleus (AGN), if any, nearby to GRB.

The possibility
of vanishing gravity and vacuum mixing angle in Eq. (\ref{deltaH}) allows us 
 to identify the
range of $\Delta f$ relevant for the neutrino magnetic moment effects 
only. Latter in this Sect., we briefly comment on the ranges of relevant 
 neutrino mixing  parameters 
for non vanishing gravity mixing angle with vanishing 
 neutrino magnetic moment.

 The case of $\bar{\nu}_{\mu}\rightarrow \nu_{\tau}$ can  be studied by
 replacing $\bar{\nu}_{e}$ with $\bar{\nu}_{\mu}$ along with corresponding 
 changes in $V_{G}$ and $\delta m^{2}$. 
 The intrinsic  flux of $\bar{\nu}_{\mu}$ may be greater than that of 
 $\bar{\nu}_{e}$ by 
a factor of $\sim $ 2 \cite{WB}, thus also possibly enhancing the expected 
 $\nu_{\tau}$ flux from GRBs through $\bar{\nu}_{\mu}\rightarrow \nu_{\tau}$. 
 However, we have checked that observationally 
 this possibility leads to quite similar results in terms of 
 event rates and are therefore not
 discussed here further. We now study in 
some detail, the various possibilities arising from relative 
comparison between $\delta $ and $V_{G}$ in Eq. (\ref{deltaH}).

	Let us first ignore the effects of VEP ($\Delta f\, =\, 0$). For 
constant $B$, the spin-flavor precession probability 
 $P(\bar{\nu}_{e}\rightarrow \nu_{\tau})$ is obtained using Eq. (\ref{deltaH})
 as
\begin{equation}
 P(\bar{\nu}_{e}\rightarrow \nu_{\tau})\, =\, 
 \left[\frac{(2\mu B)^{2}}{(2\mu B)^{2}+\delta^{2}}\right]
 \sin^{2}\left(\sqrt{(2\mu B)^{2}+\delta^{2}}
 \cdot \frac{\Delta r}{2}\right).
\label{Pdeltaf0} 
\end{equation}
We take $\mu \, \sim \, 10^{-12}\, \mu_{B}$
or less, where $\mu_{B}$ is Bohr magneton, which is less than
the stringent astrophysical upper bound on $\mu$ based on cooling of 
red giants \cite{R}. 
We here consider the transition magnetic moment, thus  
allowing the
possibility of simultaneously changing the relevant neutrino flavor as well 
 as the helicity. Therefore,  the precessed  
$\nu_{\tau}$ is an active neutrino and interacts weakly.
 In Eq. (\ref{Pdeltaf0}), $\Delta r$
 is the width of the region with $B$. If 
$\delta \, < \, 2\mu B$,
 then, for $E\, \sim \, 2\times 10^{6}$ GeV and using Eq. (\ref{bprofile}), 
 we obtain $\delta m^{2}\, < \, 
 5\times 10^{-8}$ eV$^{2}$. We take, $\delta m^{2}\, \sim \, 10^{-9}$
  eV$^{2}$, as an example and consequently we obtain from 
 Eq. (\ref{Pdeltaf0}) an energy 
 independent large ($P\, >\, 1/2$) spin-flavor 
precession probability for $\mu \, \sim \, 10^{-12}\mu_{B}$
 with $10^{-4}\, {\buildrel < \over {_{\sim}}}\, \Delta r/\mbox{pc}
 {\buildrel < \over {_{\sim}}}\,1$.
 This relatively small value of $\delta m^{2}$ is also interesting in the 
 context of sun and supernovae \cite{ss}.  Thus, for $\mu $ of the 
 order of $10^{-12}\, \mu_{B}$,
the $\nu_{\tau}$ flux may be higher than the expected one from GRBs, that is,
 $F_{\tau}/F_{e}\, > \, 10^{-5}$
 due to neutrino spin-flavor precession effects. 
 The neutrino spin-flavor precession effects are
 essentially determined by the product $\mu B$ so one may rescale  
 $\mu $ and $B$ to obtain the same results. 
 For $\delta \, \simeq 2\mu B$ and $\delta \, > \, 
 2 \mu B$, we obtain from Eq. (\ref{Pdeltaf0}), an energy dependent $P$ such
 that $P\, <\, 1/2$.

	With non vanishing $\Delta f$ ($\Delta f\, \neq \, 0$), a resonant 
character 
in neutrino spin-flavor precession can be obtained for a range of  
 values of relevant neutrino 
mixing parameters\footnote{\footnotesize\tightenlines From above discussion, 
 it follows that $E$ 
 dependent/independent spin-flavor precession  
   may also be obtained for non zero $\Delta f$, 
 however, given the current status of the high energy neutrino detection, 
 for simplicity, we ignore these possibilities which tend to 
 overlap with this case for a certain range of relevant 
 neutrino mixing parameters; for details of these possibilities in the 
 context of  AGN, see \cite{athar}.}. Two conditions are 
 essential to obtain a resonant character in
neutrino spin-flavor precession: the level crossing and the adiabaticity 
 at the level crossing (resonance).
The level crossing condition is obtained by taking $\Delta H\, =\, 0$ and 
 is given by:
\begin{equation}
 \delta m^{2}\, \sim \, 10^{-3} \mbox{eV}^{2} 
 \left(\frac{|\Delta f|}{10^{-28}}\right).
\label{levelcrossing}
\end{equation}  
These $\Delta f$ values are well below the relevant upper limits on $\Delta f$
 which are typically in the $10^{-20}$ range \cite{recent}. 
 Conversely speaking, the prospective detection of high energy neutrinos from
cosmologically distant GRBs may be sensitive to $\Delta f$ values as low as
 $\sim 10^{-28}$. The other essential condition, namely, 
 the adiabaticity in the resonance  reads \cite{adiabaticity}
\begin{equation}
 \kappa \, \equiv \, \frac{2(2\mu B)^{2}}{|\mbox{d}V_{G}/\mbox{d}r|}
 \, {\buildrel > \over {_{\sim}}}\, 1.
\label{adiabaticity}
\end{equation}
Note that here  $\kappa $ depends
explicitly on $E$ through $V_{G}$ unlike the case of ordinary neutrino 
 spin-flip induced by the matter effects. 
 A resonant character in neutrino spin-flavor precession is obtained if 
$\kappa \,  {\buildrel > \over {_{\sim}}}\, 1$ such that Eq. 
 (\ref{levelcrossing}) is satisfied.  We notice that 
$B_{ad}/B_{GRB}\, {\buildrel <\over {_{\sim}}}\, 1$ for $\mu \, 
 \sim 10^{-12}\mu_{B}$. Here $B_{ad}$ is obtained by setting 
$\kappa \, \sim \, 1$ in Eq. (\ref{adiabaticity}). 
 The general expression for relevant neutrino spin-flavor conversion 
 probability is given by \cite{ICTP} 
\begin{equation}
 P(\bar{\nu}_{e}\rightarrow \nu_{\tau})\, =\, 
 \frac{1}{2}-\left(\frac{1}{2}-P_{LZ}\right)\cos 2\theta_{f}\cos 2\theta_{i},
\label{PLZ}
\end{equation}
where $P_{LZ}\, =\, \exp(-\frac{\pi}{4}\kappa)$ and $
 \tan 2\theta_{i}\, =\, (2\mu B)/\Delta H$
 is being evaluated at the high energy neutrino 
 production site in the vicinity of GRB, whereas $\tan 2\theta_{f}\, =\, 
 (2\mu B)/\delta $ is evaluated at the exit. 
In Fig. 1, we plot $P(\bar{\nu}_{e}\rightarrow \nu_{\tau})$
 given by Eq. (\ref{PLZ}) as a function of 
$\Delta f $ as well as $\delta m^{2}$ with $E \sim 5\times 10^{6}$ GeV. 
 Four equi $P$ contours are also shown in Fig. 1. 
 Note that the resonant spin-flavor precession probability 
 is relatively small ($P < 1/2$) for $\Delta f {\buildrel > \over {_{\sim}}}
 10^{-26}$ essentially irrespective of $\delta m^{2}$ values. 
 The expected   spectrum $F_{\tau}$ of the high 
 energy tau  neutrinos originating from GRBs due to spin-flavor conversions 
 is calculated as \cite{ICTP}
\begin{equation}
 F_{\tau}\simeq P(\bar{\nu}_{e}\rightarrow \nu_{\tau})F^{0}_{e}.
\label{FTAU}
\end{equation}
 The energy dependence in $F_{\tau}$ is now evident [as compared to 
 $F_{\tau}$ given by Eq. (\ref{falpha})] when we convolve 
 $P(\bar{\nu}_{e}\rightarrow \nu_{\tau})$ given by Eq. (\ref{PLZ}) with 
 $F^{0}_{e}$ taken from Ref. \cite{WB}. The degree of energy dependence 
 clearly depends on the extent of spin-flavor conversions.   
 With the improved information on either $\Delta f$
 and/or $\mu $, one may be able to  distinguish between the situations of
 resonant and non resonant spin-flavor precession  induced by  an interplay 
 of VEP and $\mu $ in $B_{GRB}$.

	Let us now consider briefly the effects of non vanishing gravity 
 mixing angle  $\theta_{G}$  for vanishing neutrino magnetic moment. 
 In the case of massless or
degenerate neutrinos, the corresponding vacuum flavor oscillation analog 
 for $\nu_{e}\rightarrow \nu_{\tau}$ is obtained through 
 $\theta \rightarrow \theta_{G}$ and $\frac{\delta m^{2}}{4E} \rightarrow 
 V_{G}$ in the standard flavor precession probability formula in 2 flavor 
 approxiamtion. For maximal $\theta_{G}$, the sensitivity of $\Delta f$
 may be estimated by equating the argument of second $\sin$ factor equal to
$\pi/2$ in the corresponding expression for $P$ \cite{MS}. 
 This implies $\Delta f\, \sim \, 10^{-41}$ with $\phi(r)\, \simeq \, 
 \phi_{SC}(r)$. This value of $\Delta f$ is
 of the same order of magnitude as that expected for neutrinos originating 
 from AGNs. In case of non zero $\delta m^{2}$, a resonant 
 or/and non resonant flavor conversion between $\nu_{e}$ and $\nu_{\tau}$ 
 in the vicinity of a GRB is also possible due to an
 interplay of vanishing/non vanishing vacuum and gravity mixing angles. 
 For instance, 
 a resonant flavor conversion between $\nu_{e}$ and $\nu_{\tau}$ may be 
 obtained if $\sin^{2}2\theta_{G}\, \gg \, 0.25$ with $\Delta f \, \sim 
 10^{-31}$  
 ($\theta \, \rightarrow \, 0$). Here the relevant level 
crossing may occur at $r\, \sim \, 0.1 $ pc with corresponding 
 $\delta m^{2}\, \sim \, 10^{-6}$ eV$^{2}$. 

\section{Signatures of high energy $\nu_{\tau}$ in neutrino telescopes}

	The  km$^{2}$ surface area under water/ice  high energy neutrino 
 telescopes may be able to obtain first examples of high energy $\nu_{\tau}$, 
 through {\em double showers}, 
originating from GRBs correlated in time and direction with corresponding 
gamma-ray burst or may at least provide relevant useful upper limits \cite{L}. 
 The first shower occurs because of deep inelastic charged current interaction 
 of high energy 
 tau neutrinos near/inside the neutrino telescope producing the tau lepton
 (along with the first shower) 
 and the second shower occurs due to (hadronic) decay of this tau lepton.

The calculation of down ward going contained but separable  
 double shower event rate for a km$^{2}$ surface area under ice/water 
 neutrino telescope can be carried out by  replacing 
 the muon range expression with the tau one 
 ($\sim E(1-y)\tau c/m_{\tau} c^{2}$) and then subtracting  it from the
linear size of a typical high energy neutrino telescope  
 in the event rate formula while using the expected 
$\nu_{\tau}$ flux spectrum given by Eq. (\ref{falpha}) and/or by 
 Eq. (\ref{FTAU}). Here, $y$ is the fraction of the neutrino energy carried by 
 the hadrons in lab frame. Thus, $(1-y)$ is the fraction of energy 
transferred to the associated tau lepton having life time $\tau c$ and mass
 $m_{\tau}c^{2}$. We take here 
 $y \sim 0.25$ \cite{L}. 
 The condition of containdness of the two showers is obtained by requiring 
 that the 
 separation between the two showers is less than the typical $\sim $ km 
size of the neutrino telescope. It is obtained by equating the range of tau 
 neutrino induced tau leptons with the linear size of detector implying 
$E \,  {\buildrel < \over {_{\sim}}}\, 2\times 10^{7}$ GeV. 
 The condition of separableness of the two showers is 
obtained by demanding that the separation between the two showers is 
larger  than 
 the typical spread of the showers such that the amplitude of the second shower
 is essentially 2 times the first shower. This leads to 
$E \,  {\buildrel > \over {_{\sim}}}\, 2\times 10^{6}$ GeV \cite{L}. 
 Thus, the two showers may be separated by a $\mu$-like track within these 
energy limits. 
To calculate the event rates, we use Martin Roberts Stirling 
 (MRS 96 R$_{1}$) parton distributions \cite{mrs} and present event rates 
 in units of yr$^{-1}$
sr$^{-1}$. We have checked that other recent parton  distributions give quite 
 similar event rates and are therefore not depicted here.   
 Following \cite{APZ,apz}, we present in Table I, the expected 
 contained but separable  double shower event rates for down word going 
 $\nu_{\tau}$ 
in  km$^{2}$ size under water/ice  \v{C}erenkov high energy neutrino 
 telescopes for illustrative purposes. In Table I, the vacuum oscillation 
 situation is essentially independent of the choice of the oscillation 
 solution to solar neutrino problem.  
 From Table I, we notice that the event 
 rates for neutrino flavor/spin-flavor precession are up to $\sim $ 5 orders 
 of magnitude higher than that for typical intrinsic (no oscillations) tau   
 neutrino flux.

The possibility of measuring the contained but separable double shower events
 may enable one to distinguish between the high energy tau neutrinos and 
 electron
 and/or muon neutrinos originating from cosmologically distant GRBs 
 while providing useful information about 
 the relevant energy interval at the same time. 
 The chance of having  double shower  events
 induced by electron and/or muon neutrinos is negligibly small
 for the relevant energies \cite{L}. Collective information about
directionality of the source, rate and energy dependence of neutrino fluxes 
 will be needed  to possibly isolate the mechanism of neutrino oscillation.   
The up ward going 
 tau neutrinos at these energies may lead to a small pile up of 
 up ward going $\mu$-like 
 events near (10$^{4}-10^{5}$) GeV with fairly flat zenith angle 
 dependence \cite{HS}.

We now briefly discuss the potential of the under water/ice high 
energy neutrino telescopes to 
possibly determine an observational consequence of neutrino spin-flip 
in GRB induced by VEP. In the electron neutrino channel, 
 the $\bar{\nu}_{e}$ interaction rate (integrated over all angles) 
is estimated to be  an order of magnitude higher than that of 
 $(\nu_{e}+\bar{\nu}_{e})$ per
 Megaton year \cite{APZ}. This an order of magnitude  difference in 
interaction rate of {\em down ward going}   
$\bar{\nu}_{e}$ is due to Glashow resonance encountered by 
 $\bar{\nu}_{e}$ with $E\, {\buildrel > \over {_{\sim}}} \, 10^{6}$ GeV
 when $\bar{\nu}_{e}$  interact with electrons inside the detector as 
 compared to corresponding deep inelastic scattering. 
 The up ward going $\bar{\nu}_{e}$, on the other hand, while passing 
 through the earth, at these energies, are almost completely absorbed 
 by the earth mainly due to same resonant effect. 
 Thus, for instance, 
 if $E \, \sim \, 6.4\times 10^{6}$ GeV, an energy
resolution $\Delta E/E\, \sim \, 2\Gamma_{W}/M_{W}\, \sim \, 1/20$, 
where $\Gamma_{W}\, \sim $ 2 GeV is the 
width of Glashow resonance and $M_{W}\, \sim \, $80 GeV, may be needed
 to empirically differentiate between $\bar{\nu}_{e}$ and 
 $(\nu_{e}+\bar{\nu}_{e})$.  The existing/planned 
high energy neutrino telescopes may thus in principle attempt to measure the 
$\nu_{e}/\bar{\nu}_{e}$ ratio in addition to identifying 
 ($\nu_{\tau}+\bar{\nu}_{\tau}$) and ($\nu_{\mu}+\bar{\nu}_{\mu}$) 
 events  
separately. 

	This feature may be utilized, for instance, to explain a  situation 
 in which a {\em change} in $\nu_{e}/\bar{\nu}_{e}$ ratio is observed  as 
compared to GRB neutrino flux
predictions in \cite{WB}. This situation, if realized obsevationally 
 may  be an evidence for the neutrino spin-flip in GRB due to VEP, provided if 
 neutrinos and antineutrinos couple differently to gravity.
  This follows from the possibility  discussed in previous Sect. 
 that an interplay between VEP and neutrino magnetic moment in $B_{GRB}$ may 
 leads to conversions in either $\nu_{e}$ or $\bar{\nu}_{e}$ 
channel but not in both channels simultaneously.

\section{Results and discussion}

		1. Intrinsically, the flux of high energy cosmic tau 
 neutrinos is quite
small, relative to non tau neutrino flavor, typically being 
 $F^{0}_{\tau}/F^{0}_{e, \mu}\, <\, 10^{-5}$ (whereas 
 $F^{0}_{e}/F^{0}_{\mu} \sim 1/2$) from cosmologically distant GRBs.

	2. Because of neutrino oscillations, this ratio can be greatly 
 enhanced. In
the context of three flavor neutrino mixing scheme which can accommodate the 
oscillation solutions to solar and atmospheric neutrino deficits in terms of 
 oscillations between three active neutrinos, the final down ward going
ratio of fluxes of high energy cosmic neutrinos on earth is 
 $F_{e}\sim F_{\mu}\sim F_{\tau} \sim
F^{0}_{e}$, essentially irrespective of the oscillation solution to solar 
neutrino problem.

The (vacuum) flavor oscillations leads to an essentially energy 
 independent
flux of high energy neutrinos of all flavors originating from cosmologically 
 distant GRBs at the level of electron neutrino
flux, whereas spin-flavor precessions/conversions may lead to an energy 
 dependence or/and change in this situation.  

The spin-flavor conversions may occur possibly through several
 mechanisms. We have discussed in some detail mainly the spin-flavor
 precession/conversion situation induced by a non zero neutrino magnetic moment
 and by a relatively small VEP as an example  
 to point out the possibility of obtaining some what higher tau neutrino 
 fluxes as compared
 to no oscillations/conversions scenarios from GRBs.     

	The matter density in the vicinity of GRB
is quite small (up to 4$-$5 orders of magnitude) to induce any resonant 
 flavor/spin-flavor neutrino 
 conversion due to normal matter effects.
We have pointed out that a resonant character in the neutrino spin-flavor 
 conversions 
may nevertheless be obtained due to possible VEP. The 
corresponding degree of VEP  may be
 $\sim \, (10^{-35}-10^{-25}$) depending on $\delta m^{2}$ value
 for vanishing gravity mixing angle.

	3. This enhancement in high energy cosmic tau neutrino flux may lead 
to the possibility of its detection in km$^{2}$ surface area high energy 
neutrino telescopes. For $2\times 10^{6}\leq E$/GeV
$\leq 2\times 10^{7}$, the down ward going high energy cosmic tau neutrinos
may produce a double shower signature because of charged current deep 
inelastic scattering followed by a subsequent hadronic decay of associated tau
lepton

	The double shower  event rate for intrinsic 
 (no oscillations/conversions) high energy
 tau neutrinos originating from GRBs turns out to be small as compared to 
 that due to precession/conversion effects up to a factor of 
 $\sim \, 10^{-5}$. Thus, the high energy neutrino telescopes may possibly 
 provide useful upper bounds on intrinsic properties of neutrinos such as
 mass, mixing and magnetic moment, etc.. The relevant tau neutrino energy 
 range for detection in km$^{2}$ surface area under water/ice neutrino 
 telescopes may be  $2\times 10^{6}
 {\buildrel < \over {_{\sim}}}\, E/\mbox {GeV}\,
{\buildrel < \over {_{\sim}}} \, 2\times 10^{7}$ through characteristic 
 contained but separable double shower  events. 

Observationally, the high energy $\nu_{\tau}$ burst from a GRB may possibly  
be {\em correlated} to the corresponding gamma-ray burst/highest energy cosmic 
 rays (if both have common origin) in time and in 
direction thus raising the possibility of its detection. 
 If the range of neutrino mixing parameters pointed out in this 
study is realized terrestially/extraterrestially then a relatively large 
 (energy dependent) $\nu_{\tau}$ flux from GRBs is 
expected as compared to no oscillation/conversion scenario.

\paragraph*{Acknowledgments.} 
 
The author thanks Japan Society for the Promotion of Science 
for financial support.

\pagebreak

\begin{table}[th]
\caption{Event rate (yr$^{-1}$sr$^{-1}$) for down word going high energy tau 
 neutrino induced contained but separable double
 showers connected by a $\mu-$ like track in various energy bins using 
 MRS 96 R$_{1}$ parton distributions.
 For spin-flavor precessions, we use $\delta m^{2}\, 
 {\buildrel < \over {_{\sim}}}\, 10^{-9}$ eV$^{2}$ and $\mu \, \sim 
 10^{-12}\mu_{B}$ [see Eq. (\ref{Pdeltaf0}) in the text], 
 whereas for vacuum flavor
 oscillations, we used Eq. (\ref{falpha}).}
\begin{tabular}{cccc}  
 {\em Energy Interval} & 
 \multicolumn{3}{c}{\em Rate (yr$^{-1}$sr$^{-1}$)} \\ \cline{2-4}
    & {\em no  osc} & {\em vac osc} 
 & {\em spin-flavor precession} \\ \cline{1-4}
 $2\times 10^{6}{\buildrel < \over {_{\sim}}}\, E/\mbox {GeV}\,
{\buildrel < \over {_{\sim}}} \, 5\times 10^{6}$  &$10^{-6}$ &
 $1\times 10^{-1}$ & $0.5\times 10^{-1}$   \\ 
 $5\times 10^{6}{\buildrel < \over {_{\sim}}}\, E/\mbox {GeV}\,
{\buildrel < \over {_{\sim}}} \, 7\times 10^{6}$  &  $2\times 10^{-7}$ &
 $2\times 10^{-2}$  & $10^{-2}$  \\ 
  $7\times 10^{6}{\buildrel < \over {_{\sim}}}\, E/\mbox {GeV}\,
{\buildrel < \over {_{\sim}}} \, 1\times 10^{7}$   & $2\times 10^{-7}$ &
 $2\times 10^{-2}$   & $10^{-2}$ \\
  $1\times 10^{7}{\buildrel < \over {_{\sim}}}\, E/\mbox {GeV}\,
{\buildrel < \over {_{\sim}}} \, 2\times 10^{7}$   & $2\times 10^{-7}$ & 
 $2\times 10^{-2}$   & $10^{-2}$ 
\end{tabular}
\end{table}

\pagebreak

\begin{figure}[t]
\leavevmode
\epsfxsize=5in
\epsfysize=5in
\epsfbox{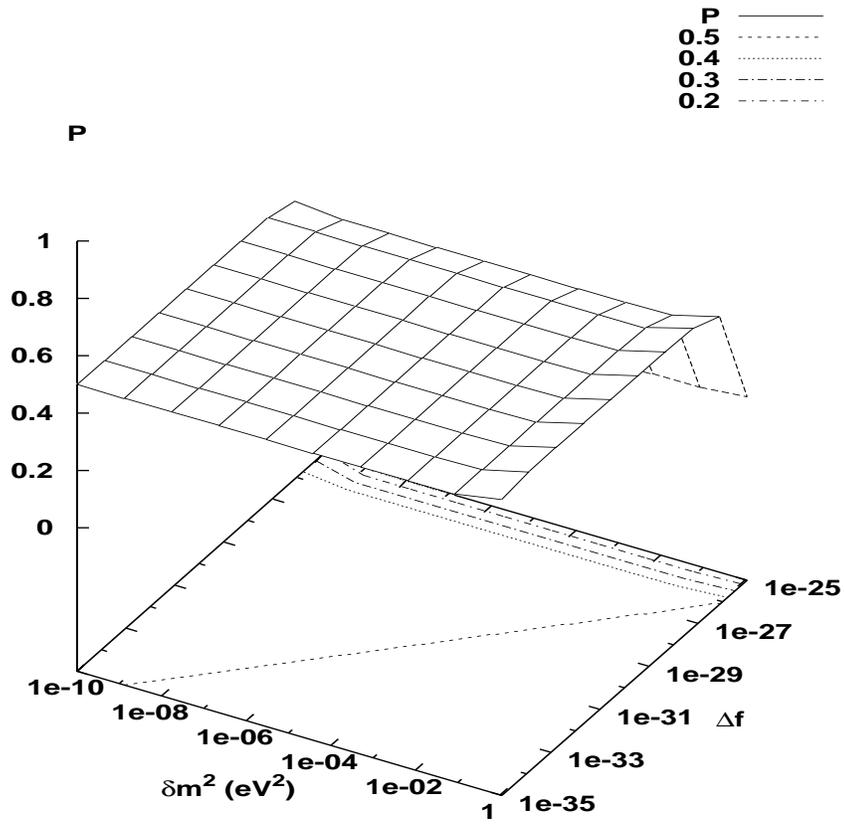}
\caption{$P(\bar{\nu}_{e}\rightarrow \nu_{\tau})$ using Eq. (\ref{PLZ}) as a 
 function of $\delta m^{2}$ (eV$^{2}$)  and $\Delta f$.}
\end{figure}

\end{document}